\documentclass[nofootinbib,aps,11pt]{revtex4-1}
\usepackage{graphicx}
\usepackage{hyperref}
\usepackage{cancel}
\usepackage{amssymb}
\usepackage{textcomp}
\usepackage{amsmath}
\usepackage{bm}
\usepackage{times}
\usepackage{epsfig}
\usepackage{color}
\usepackage{feynmp}
\newcommand{\eV}{{\rm eV}}

\newcommand{\GeV}{{\rm GeV}}
\newcommand{\TeV}{{\rm TeV}}

\begin{document}
\title{\LARGE Interpreting the $R_{K^{(*)}}$ Anomaly in the Colored Zee-Babu Model}
\bigskip

\author{Shu-Yuan Guo$^{1}$}
\email{shyuanguo@mail.nankai.edu.cn}
\author{Zhi-Long Han$^{2}$}
\email{sps\_hanzl@ujn.edu.cn}
\author{Bin Li$^{1}$}
\email{Now at School of Physics, Peking University, libin@pku.edu.cn}
\author{Yi Liao~$^{1,3,4}$}
\email{liaoy@nankai.edu.cn}
\author{Xiao-Dong Ma~$^{1}$}
\email{maxid@mail.nankai.edu.cn}

\affiliation{
$^1$~School of Physics, Nankai University, Tianjin 300071, China
\\
$^2$ School of Physics and Technology, University of Jinan, Jinan, Shandong 250022, China
\\
$^3$ CAS Key Laboratory of Theoretical Physics, Institute of Theoretical Physics,
Chinese Academy of Sciences, Beijing 100190, China
\\
$^4$ Center for High Energy Physics, Peking University, Beijing 100871, China
}
\date{\today}

\begin{abstract}
We consider the feasibility of interpreting the $R_{K^{(*)}}$ anomaly in the colored Zee-Babu model. The model generates neutrino masses at two loops with the help of a scalar leptoquark $S\sim(3,3,-\frac{1}{3})$ and a scalar diquark $\omega\sim(6,1,-\frac{2}{3})$, and contributes to the transition $b\to s\ell^-\ell^+$ via the exchange of a leptoquark $S$ at tree level. Under constraints from lepton flavor violating (LFV) and flavor changing neutral current (FCNC) processes, and direct collider searches for heavy particles, we acquire certain parameter space that can accommodate the $R_{K^{(*)}}$ anomaly for both normal (NH) and inverted (IH) hierarchies of neutrino masses. We further examine the LFV decays of the $B$ meson, and find a strong correlation with the neutrino mass hierarchy, i.e., $\text{Br}(B^+ \to K^+ \mu^\pm\tau^\mp)\gtrsim\text{Br}(B^+ \to K^+ \mu^\pm e^\mp)
\approx\text{Br}(B^+ \to K^+ \tau^\pm e^\mp)$ for NH, while
$\text{Br}(B^+ \to K^+ \mu^\pm \tau^\mp)\ll\text{Br}(B^+ \to K^+ \mu^\pm e^\mp)
\approx\text{Br}(B^+ \to K^+ \tau^\pm e^\mp)$ for IH. Among these decays, only $B^+ \to K^+ \mu^\pm e^\mp$ in the case of NH is promising at the LHCb RUN II, while for IH all LFV decays are hard to detect in the near future.
\end{abstract}

\maketitle

\section{Introduction}

Neutrino oscillation experiments suggest that neutrinos have masses, which provides a definite piece of evidence for physics beyond the standard model (SM). Yet, the origin of tiny neutrino masses still remains unknown. Many authors~\cite{Ma:1998dn,Ma:2000cc,Krauss:2002px,Ma:2006km,Ma:2007gq,Liao:2010ku,Liao:2010cc,
Chen:2011de,McDonald:2013kca,Boucenna:2014zba,Wang:2015saa,Wang:2016vfj,Ma:2016mwh,Megrelidze:2016fcs,
Wang:2016lve,Wang:2017mcy,Cai:2017jrq} propose that neutrino masses might originate from physics at the $\TeV$ scale, whose effects could be tested at the LHC~\cite{Han:2006ip,delAguila:2007qnc,Franceschini:2008pz,Perez:2008ha,
delAguila:2008cj,Ding:2014nga,Han:2015hba,Deppisch:2015qwa,Han:2015sca,Ding:2016wbd,Guo:2016dzl,
Guo:2017ybk}. Recently, the LHCb collaboration has found hints for the violation of lepton flavor universality in the semi-leptonic decays of the $B$ meson by measuring the ratio $R_{K^*}=\text{Br}(B\to K^* \mu^+\mu^-)/\text{Br}(B\to K^* e^+e^-)$~\cite{Aaij:2017vbb}:
\begin{eqnarray}
R_{K^*}=\left\{
\begin{array}{cl}
0.66^{+0.11}_{-0.07}\pm 0.03 & \text{for } 0.045<q^2<1.1~\GeV^2, \\
0.69^{+0.11}_{-0.07}\pm 0.05 & \text{for } 1.1<q^2<6.0~\GeV^2,
\end{array} \right.
\end{eqnarray}
which differs from the SM prediction~\cite{Hiller:2003js} by $2.1-2.3\sigma$ in the low $q^2$ region and $2.4-2.5\sigma$ in the medium $q^2$ region. Here $q^2$ is the momentum squared of the leptonic system. A few years earlier, the LHCb also measured the ratio~\cite{Aaij:2014ora},
\begin{equation}
R_K =\frac{\text{Br}(B\to K \mu^+\mu^-)}{\text{Br}(B\to K e^+e^-)}
=0.745^{+0.090}_{-0.074}\pm0.036,
\end{equation}
for $1<q^2<6~\GeV^2$, which is $2.6\sigma$ deviation from the SM prediction~\cite{Bordone:2016gaq}.
In addition, ATLAS~\cite{ATLAS:2017dlm}, Belle~\cite{Wehle:2016yoi}, and LHCb~\cite{Aaij:2015oid} observed the so-called $P_5^\prime$ angular observable discrepancy in the decay $B\to K^\ast \mu^+\mu^-$, respectively with a significance of $2.7\sigma$ for $q^2\in[4,6]~\GeV^2$, $2.6\sigma$ and $3.4\sigma$ for $q^2\in[4,8]~\GeV^2$.
But the CMS~\cite{CMS:2017ivg} measurement in the same decay channel is consistent with the SM. In the meanwhile, LHCb also found a $3.5\sigma$ disagreement with the SM prediction in the measurement of differential branching fraction and angular analysis of the decay $B_s^0\to\phi\mu^+\mu^-$. All these anomalies involve the same parton level process $b\to s\ell\ell~(\ell=e,~\mu)$, hinting at potential new physics beyond the SM, and have caused vigorous discussions on their possible origin~\cite{Altmannshofer:2013foa,Altmannshofer:2014cfa,Alonso:2014csa,
Hiller:2014yaa,Ghosh:2014awa,Hurth:2014vma,Glashow:2014iga,Altmannshofer:2014rta,Gripaios:2014tna,
Jager:2014rwa,Bhattacharya:2014wla,Crivellin:2015mga,Crivellin:2015lwa,Crivellin:2015era,Alonso:2015sja,
Calibbi:2015kma,Descotes-Genon:2015uva,Bauer:2015knc,Ciuchini:2015qxb,Hurth:2016fbr,Boucenna:2016wpr,
Das:2016vkr,Feruglio:2016gvd,Boucenna:2016qad,Becirevic:2016oho,Becirevic:2016yqi,Altmannshofer:2016jzy,
Sahoo:2016pet,Hiller:2016kry,Bhattacharya:2016mcc,Capdevila:2017ert,Ko:2017yrd,Altmannshofer:2017fio,
Crivellin:2017zlb,Capdevila:2017bsm,Altmannshofer:2017yso,Hiller:2017bzc,Geng:2017svp,Ciuchini:2017mik,
DAmico:2017mtc,Ghosh:2017ber,Alok:2017jaf,Alok:2017sui,Feruglio:2017rjo,Tang:2017gkz,Hurth:2017hxg,Datta:2017ezo,
Bardhan:2017xcc,Das:2017kfo,DiLuzio:2017chi,Chauhan:2017ndd,King:2017anf,Dorsner:2017ufx,
Choudhury:2017qyt,Crivellin:2017dsk}.

In this paper we explore the possibility that both neutrino masses and the $R_{K^{(*)}}$ anomaly have a common origin; see Refs.~\cite{Boucenna:2015raa,Pas:2015hca,Deppisch:2016qqd,Cheung:2016fjo,Cheung:2016frv,
Cheung:2017efc,Ko:2017quv,Bhatia:2017tgo,Dorsner:2017wwn,Cai:2017wry,Chiang:2017hlj,He:2017osj} for some of earlier discussions in this spirit. While previous studies pay less attention to the inverted neutrino mass hierarchy (IH) than the normal hierarchy (NH), we stress that it is worthwhile to examine both in detail. To be specific, we consider the colored Zee-Babu model proposed in Ref.~\cite{Babu:2001ex} to interpret the $R_{K^{(*)}}$ anomaly. Some other possible phenomenologies in this model have been explored in Refs.~\cite{Kohda:2012sr,Nomura:2016ask,Chang:2016zll}. By adopting a suitable parametrization, we find that the model with neutrino masses in either NH or IH could accommodate the $R_{K^{(*)}}$ anomaly. More interestingly, the model predicts measurable lepton flavor violating (LFV) decays of the $B$-meson at the LHCb RUN II~\cite{Boucenna:2015raa}, in particular the decay $B\to K e^\pm \mu^\mp$ in the case of NH, which could be used to distinguish between the two hierarchies.

The rest of this paper is organised as follows. In Sec.~\ref{MC} the colored Zee-Babu model and constraints on it are briefly reviewed. In Sec.~\ref{RK}, we perform a scan of parameter space for both NH and IH to interpret the $R_{K^{(*)}}$ anomaly, and discuss LFV decays of the $B$-meson. Finally, our conclusions are summarized in Sec.~\ref{CL}.

\section{Model and Constraints}\label{MC}

\subsection{Model Setup}

\begin{figure}
	\centering
	\includegraphics[width=0.45\linewidth]{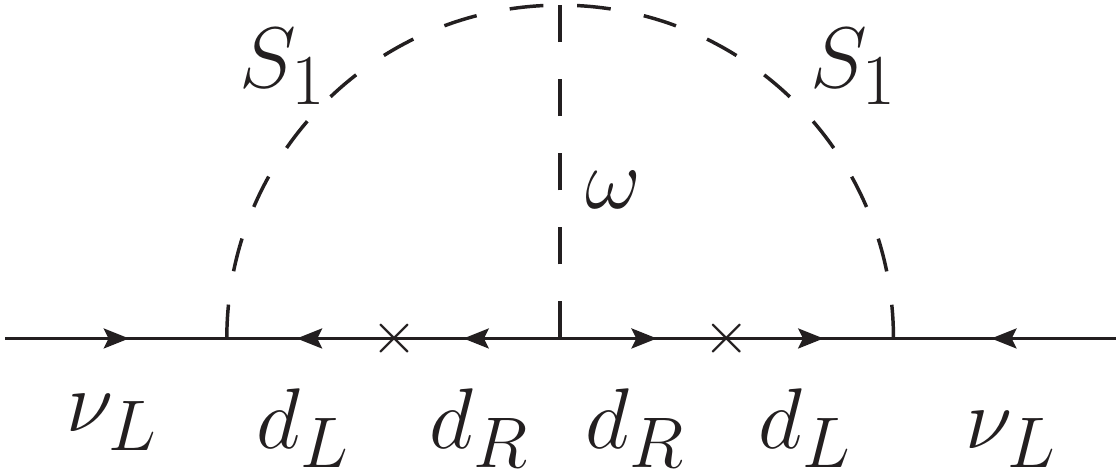}
\caption{Neutrino masses generated at two loops.}
\label{Fig:neutrinomass}
\end{figure}

The colored Zee-Babu model~\cite{Babu:2001ex} introduces a scalar leptoquark $S\sim(3,3,-\frac{1}{3})$ and a scalar diquark $\omega\sim(6,1,-\frac{2}{3})$ under the SM gauge group $SU(3)_C\times SU(2)_L\times U(1)_Y$. Expressing $S$ as a matrix in $SU(2)_L$ and $\omega$ as a matrix in $SU(3)_C$,
\begin{equation}
	S =\frac{1}{\sqrt{2}}
	\begin{pmatrix}
	S_1          & \sqrt{2}S_3 \\
	\sqrt{2}S_2  & -S_1
	\end{pmatrix},~ \quad
	\omega=\frac{1}{\sqrt{2}}
	\begin{pmatrix}
	\sqrt{2}\omega_1 & \omega_4 & \omega_5 \\
	\omega_4 & \sqrt{2}\omega_2 & \omega_6 \\
	\omega_5 & \omega_6 & \sqrt{2}\omega_3
	\end{pmatrix},
\end{equation}
the Yukawa couplings relevant for neutrino masses are
\begin{align}
	- \mathcal{L}_Y =~&y_S^{ij} \overline{(L_{Li})^C} i\sigma_2 {S^{*\alpha}} Q_{L j}^{\alpha} + y_\omega^{ij} \overline{(d_{Ri}^{\alpha})^C} d_{Rj}^{\beta} \omega^{* \alpha \beta} + \text{h.c.},
\end{align}
where $i, j$ are the flavor indices and $\alpha,\beta$ the color indices. The diquark Yukawa matrix $y_\omega$ is obviously symmetric while $y_S$ for the leptoquark is generally complex. As shown in Fig.~\ref{Fig:neutrinomass}, the neutrino masses are generated at two-loop level with the help of the above Yukawa couplings and the trilinear coupling between the leptoquark and diquark contained in the complete scalar potential:
{\begin{align}
	V = &-\mu_\Phi^2 \Phi^\dagger \Phi + \mu_S^2 \text{Tr}(S^\dagger S) + \mu_\omega^2 \text{Tr}(\omega^\dagger \omega) +\big[\mu \text{Tr}(S \omega^* S) +\text{h.c.}\big]\notag \\
	&+ \lambda_\Phi (\Phi^\dagger \Phi)^2 + \lambda_{S1} [\text{Tr}(S^\dagger S)]^2 +  \lambda_{S2} \text{Tr}(S^\dagger S S^\dagger S) + \lambda_{S3} \text{Tr}(S^\dagger i \sigma_2 S^* S^\text{T} i \sigma_2 S)  \\
	&+ \lambda_{\omega 1} [\text{Tr}(\omega^\dagger \omega)]^2 + \lambda_{\omega 2} \text{Tr}(\omega^\dagger \omega \omega^\dagger \omega) + \lambda_{\Phi S1}(\Phi^\dagger \Phi) \text{Tr}(S^\dagger S) + \lambda_{\Phi S 2} (\Phi^\dagger S S^\dagger \Phi)   \notag \\
	&+ \lambda_{\Phi \omega}(\Phi^\dagger \Phi) \text{Tr}(\omega^\dagger \omega) + \lambda_{S \omega 1} \text{Tr}(S^\dagger S) \text{Tr}(\omega^\dagger \omega) + \lambda_{S \omega 2} \text{Tr}(S^\dagger \omega \omega^\dagger S). \notag
	\end{align}
Here the trace is understood to be done separately for the weak isospin and color indices; e.g., the trace in the $\mu$ term is $S^\alpha_{ij}\omega^{*\alpha\beta}S^\beta_{ji}$ and that in the $\lambda_{S \omega 2}$ term is $S^{*\alpha}_{ij}\omega^{\alpha\beta}\omega^{*\beta\gamma}S^{\gamma}_{ij}$. When the SM Higgs doublet $\Phi$ develops the vacuum expectation value $v/\sqrt{2}$, the $S$ particles of various electric charges are generally split in mass due to their quartic couplings with $\Phi$:
\begin{align}
	m_{S_1}^2 &= \mu_S^2 + \frac{1}{4}(2\lambda_{\Phi S1}+ \lambda_{\Phi S2}) v^2, \\
	m_{S_2}^2 &= \mu_S^2 + \frac{1}{2}(\lambda_{\Phi S1} + \lambda_{\Phi S2}) v^2, \\
	m_{S_3}^2 &= \mu_S^2 + \frac{1}{2} \lambda_{\Phi S1} v^2.
\end{align}
To avoid constraints from the oblique parameters~\cite{Cheung:2016frv}, we assume these scalars are degenerate by taking $\lambda_{\Phi S1}=\lambda_{\Phi S2}=0$, and denote their mass as $m_S$. In our later numerical study, we will consider $m_{S}>1~\TeV$ to respect the direct search bounds at LHC~\cite{Khachatryan:2015vaa,Khachatryan:2016jqo,Aad:2011ch,ATLAS:2012aq,ATLAS:2013oea,
Aad:2015caa,Aaboud:2016qeg} and for the diquark $\omega$ we take $m_\omega>7~\TeV$ to escape the dijet bounds~\cite{Khachatryan:2015dcf,Aaboud:2017yvp}.

The neutrino mass matrix generated through Fig.~\ref{Fig:neutrinomass} is~\cite{AristizabalSierra:2006gb},
\begin{equation}
M_{\nu}^{\ell \ell^\prime} = 12 \mu y_S^{\ell i} m_d^i y_{\omega}^{i j} I^{i j} m_d^j y_S^{\ell^\prime j},
\end{equation}
where $\ell,~\ell^{\prime}$ denote the neutrino flavor and $i,~j$ the down-type quark flavor summed over, and $I^{ij}$ is a loop integral, which in the limit of large leptoquark and diquark masses simplifies to
\begin{equation}
I^{ij}\simeq 
\frac{1}{(4 \pi)^4} \frac{1}{m_S^2} \tilde{I}(m_{\omega}^2/m_{S}^2),
\end{equation}
with
\begin{equation}
\tilde{I}(r)=\int_0^1dx\int_0^{1-x}dy \frac{1}{x+y(y+r-1)}
\ln\left(\frac{x+ry}{y(1-y)}\right).
\end{equation}
$M_\nu$ is then diagonalized by the PMNS matrix through
\begin{equation}
m_\nu = V^\dagger M_\nu V^*.
\end{equation}
Typically, a neutrino mass $m_\nu\sim0.01~\eV$ can be realised with $\mu\sim1~\TeV$, $y_S\sim y_\omega\sim0.01$ at $m_b=4.7~\GeV$, $m_S=1~\TeV$ and $m_\omega=7~\TeV$. Considering the radiative contributions to $m_{S}$ and $m_\omega$ through the trilinear $\mu$ term, the value of $\mu\sim 1~\TeV$ also satisfies the perturbative requirement $\mu\lesssim 5\min(m_S,m_\omega)$ for $m_S\sim 1~\TeV$ and $m_\omega\sim 7~\TeV$~\cite{Babu:2002uu,Nebot:2007bc}.

\subsection{Constraints}\label{CT}

There are precision low-energy measurements that can be applied to constrain the parameters in the model under consideration. Since the new particles are heavy, it is convenient to work with an effective field theory in which their low-energy effects are represented by a series of effective interactions. For instance, the leptoquark Yukawa couplings induce lepton or quark flavor violating transitions at the tree level. These transitions can be organized in terms of the four-Fermi effective interactions involving two leptons and two quarks~\cite{Carpentier:2010ue},
\begin{equation}
\frac{1}{2m_S^2}\sum_{ijkn}C_{ijkn}\mathcal{O}^{ijkn}
=\sqrt{2}G_F\sum_{ijkn}\epsilon^{ijkn}\mathcal{O}^{ijkn}.
\end{equation}
Here $\mathcal{O}^{ijkn}$ refer to operators like
$(\overline{\ell_i}\gamma^\mu P_L \ell_j)(\overline{q_k} \gamma_\mu P_L q_n)$ and $(\overline{\nu_i}\gamma^\mu P_L \nu_j)(\overline{q_k} \gamma_\mu P_L q_n)$ after Fierz transformation, and $\epsilon^{ijkn}$ are normalized dimensionless Wilson coefficients bounded by experimental data. For the model under consideration, the most important bounds are extracted from LFV and FCNC processes. The $\mu-e$ conversion in nuclei, due to operators of the form $(\overline{\ell_i}\gamma^\mu P_L\ell_j)
(\overline{u_k}\gamma_\mu P_L u_n)$ with $i\ne j$, sets a bound on $\epsilon^{1211}$~\cite{Carpentier:2010ue}:
\begin{equation}
\big|\epsilon^{1211}\big|
=\left|\frac{y_S^{11} y_S^{21*}}{8\sqrt{2} G_F m_S^2}\right| <8.5\times 10^{-7}.
\end{equation}
And the constraint on the decay $K\to \pi \bar{\nu}\nu$ from operators like $(\overline{\nu_i}\gamma^\mu P_L \nu_j)(\overline{d} \gamma_\mu P_L s)$, sets another bound~\cite{Carpentier:2010ue}:
\begin{eqnarray}
\big|\epsilon^{ij 12}\big|
=\left|\frac{y_S^{i1}y_S^{j2*}}{8\sqrt{2}G_Fm_S^2}\right|<9.4\times 10^{-6},
\end{eqnarray}
while the bound from the analogous decay $B\to K\bar{\nu}\nu$ due to the operators $(\overline{\nu_i}\gamma^\mu P_L \nu_j)(\overline{b} \gamma_\mu P_L s)$ is much looser~\cite{Carpentier:2010ue}:
	\begin{eqnarray}
	\big|\epsilon^{ij 32}\big|
	=\left|\frac{y_S^{i3}y_S^{j2*}}{8\sqrt{2}G_Fm_S^2}\right|<1.0\times 10^{-3}.
	\end{eqnarray}

The LFV radiative decays $\ell_i \to \ell_j \gamma$ are induced at one loop by the exchange of a leptoquark $S$  with the branching ratio~\cite{Cheung:2016frv}:
\begin{equation}
\text{BR}(\ell_i\to\ell_j \gamma) \simeq \text{BR}(\ell_i\to \ell_j \bar{\nu}_j \nu_i)\frac{{3}\alpha}{256 \pi G_F^2}\frac{1}{4m_S^4}\big|(y_Sy_S^\dagger)^{ij}\big|^2.
\end{equation}
The current experimental bounds are, $\text{BR}(\mu\to e\gamma)<4.2\times10^{-13}$~\cite{TheMEG:2016wtm}, $\text{BR}(\tau\to \mu\gamma)<4.4\times10^{-8}$~\cite{Aubert:2009ag},
and $\text{BR}(\tau\to e\gamma)<3.3\times10^{-8}$~\cite{Aubert:2009ag}, which can be used to constrain $|(y_Sy_S^\dagger)^{ij}|$ for a given $m_S$. The anomalous magnetic moment of the charged lepton can also be obtained~\cite{Cheung:2016frv}:
\begin{eqnarray}
\Delta a_{\ell_i}\simeq-\frac{ 3 }{64\pi^2}\frac{m_{\ell_i}^2}{m_{S}^2}\sum_{j}|(y_{S}^{ij})|^2.
\label{Eq. muong-2}
\end{eqnarray}
Even if assuming $\sum_{j}|(y_{S}^{ij})|^2\sim 1$ and $m_S\sim1~\TeV$, one can only get $|\Delta a_\mu|\sim 5.2\times 10^{-11}$, which is far below the current limit~\cite{Bennett:2006fi}.

Finally, the diquark Yukawa couplings contribute at tree level to the neutral meson mixing of $K^0$-$\overline{K^0}$, $B_d^0$-$\overline{B_d^0}$, and $B_s^0$-$\overline{B_s^0}$. The bounds on the corresponding Wilson coefficients were obtained in Ref.~\cite{Bona:2007vi}:
\begin{eqnarray}
|\eta_K|&=&\left|\frac{y_\omega^{11} y_\omega^{22*}}{4\sqrt{2}G_F m_\omega^2}\right|<2.9\times 10^{-8},
\\
|\eta_{B_d}|&=&\left|\frac{y_\omega^{11} y_\omega^{33*}}{4\sqrt{2}G_F m_\omega^2}\right|<7.0\times 10^{-7},
\\
|\eta_{B_s}|&=&\left|\frac{y_\omega^{22} y_\omega^{33*}}{4\sqrt{2}G_F m_\omega^2}\right|<3.3\times 10^{-5}.
\end{eqnarray}

\section{The $R_{K^{(*)}}$ Anomaly}\label{RK}

In the colored Zee-Babu model the exchange of a leptoquark at tree level contributes to the decay $b \to s \ell^+ \ell^-$, and could thus be relevant to the $R_{K^{(*)}}$ anomaly indicating lepton universality violation. The general effective Hamiltonian describing the decay is parameterized as~\cite{Hiller:2014yaa},
\begin{equation}
	\mathcal{H}_{\text{eff}} = - \frac{4 G_F}{\sqrt{2}} \frac{\alpha}{4 \pi} V_{tb} V_{ts}^* \sum_i C_i(\mu) \mathcal{O}_i(\mu) + \text{h.c.},
\end{equation}
where $\mathcal{O}_i(\mu)$ are effective operators with Wilson coefficients $C_i(\mu)$ renormalized at the scale $\mu$. In the SM the dominant contributions come from $C_9^{\text{SM}}$ and $C_{10}^{\text{SM}}$ with $C_9^{\text{SM}}=-C_{10}^{\text{SM}}$ while the dipole term is negligibly small with $|C_7^{\text{SM}}|\approx 0.07|C_9^{\text{SM}}|$~\cite{Ghosh:2014awa,Altmannshofer:2013foa,Descotes-Genon:2013wba}. For the model under consideration, only the operators $\mathcal{O}_9^{\ell_i}=(\bar{s}\gamma^\mu P_L b)(\bar{\ell}_i\gamma^\mu \ell_i)$ and $\mathcal{O}_{10}^{\ell_i}=(\bar{s}\gamma^\mu P_L b)(\bar{\ell}_i\gamma^\mu\gamma_5 \ell_i)$ are induced with the Wilson coefficients
\begin{equation}
C_9^{\ell_i} = -C_{10}^{\ell_i} =
-\frac{\sqrt{2}\pi}{4\alpha G_Fm_{S}^2}
\frac{(y_{S}^{ i3})(y_{S}^{i2*})}{V_{tb}V_{ts}^*},
\end{equation}
upon applying a Fierz identity. Differently from most previous studies where the new physics contribution to $C_{9,10}^e$ is ignored, we found for the model under consideration that it can be comparable to that of $C_{9,10}^\mu$. A model independent analysis on the above operators~\cite{Hiller:2017bzc} from the $R_{K^{(*)}}$ data suggests a favored range of $0.8\lesssim\text{Re}(X^e - X^\mu)\lesssim 1.4$ where $X^{\ell_i} = C_9^{\ell_i} - C_{10}^{\ell_i}$, which translates to $0.4\lesssim C_9^e - C_9^\mu \lesssim 0.7$ for the model under study.

\begin{figure}
	\centering
	\includegraphics[width=0.85\linewidth]{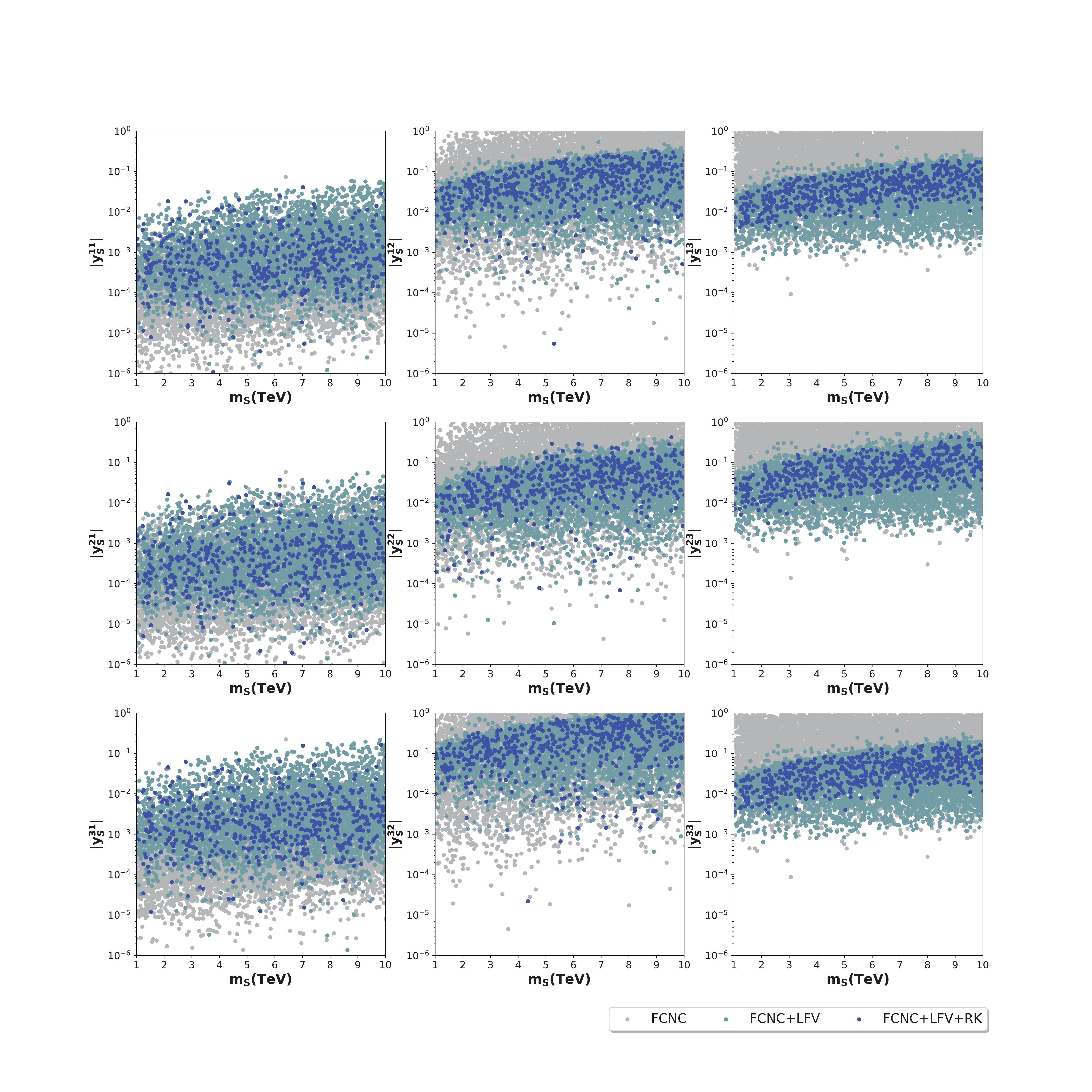}
	\caption{The relations between the leptoquark Yukawa couplings $y_S^{ij}$ and mass $m_S$ are shown for survived sample points in the case of NH.}
	\label{Fig:yuknh}
\end{figure}

\begin{figure}
	\centering
	\includegraphics[width=0.85\linewidth]{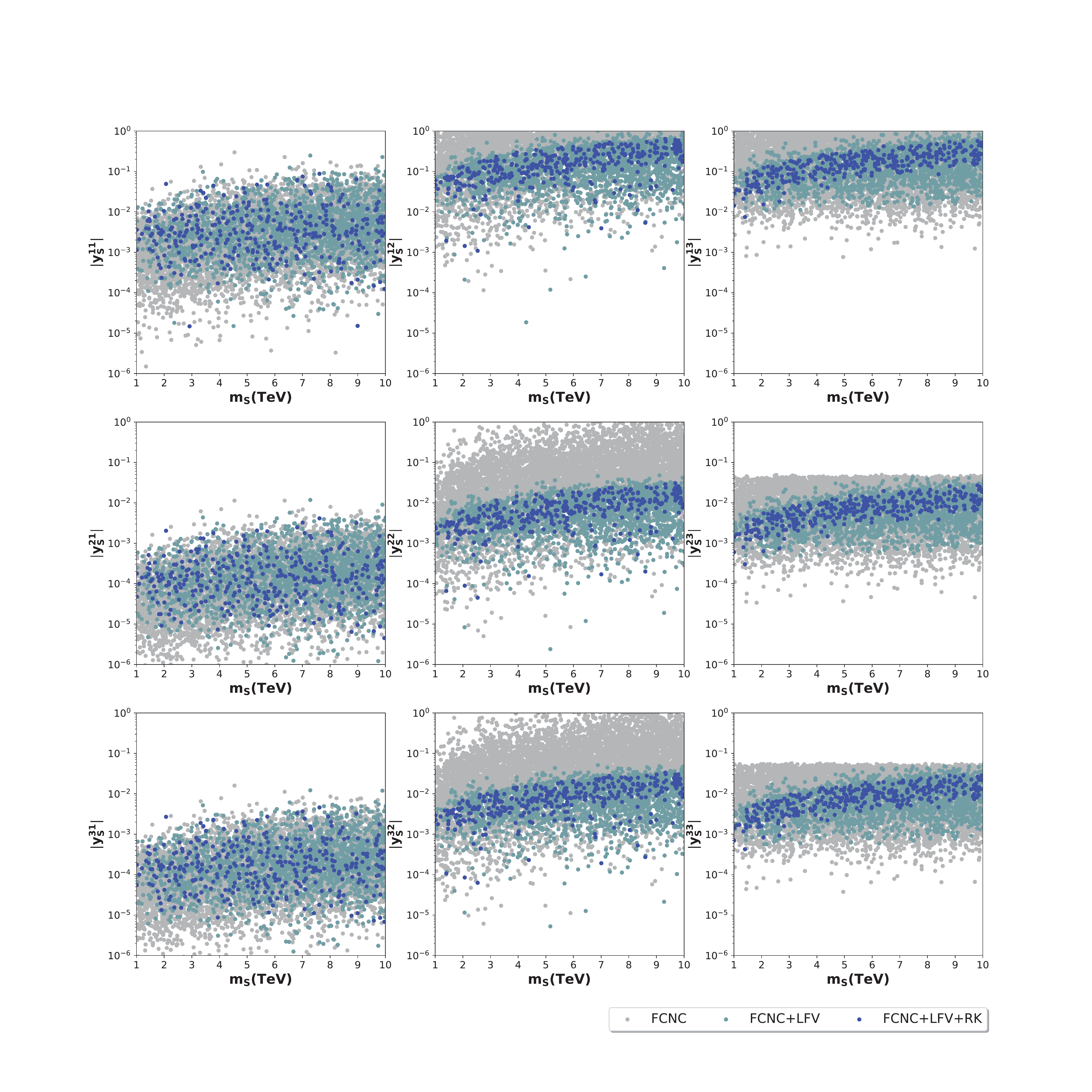}
	\caption{Same as Fig~\ref{Fig:yuknh} but for IH.}
	\label{Fig:yukih}
\end{figure}

In our numerical calculation, we work with a parametrization of the Yukawa coupling $y_S$ suggested in Ref.~\cite{Chang:2016zll} and employed the 2$\sigma$ ranges for the neutrino mixing parameters in Ref.~\cite{Capozzi:2016rtj}. For the remaining free parameters we scan in their following ranges:
\begin{equation}
\begin{aligned}
&m_1~(m_3)\in [10^{-7},10^{-3}]~\text{eV};~\mu\in [0.1,1]~\text{TeV},
~m_{S}\in [1,10]~\text{TeV},~m_{\omega}\in [7,20]~\TeV;
\\
&y_S^{13},~y_\omega^{dd},~y_\omega^{ss},~y_\omega^{sb},~y_\omega^{bb}\in [10^{-5},1];
~y_\omega^{ds}\in [0,1],~y_\omega^{db}\in [0,3];
\end{aligned}
\end{equation}
where $m_1~(m_3)$ is the lightest neutrino mass in the case of NH (IH). We apply the constraints from FCNC, LFV, and $R_{K^{(*)}}$ sequentially. In Figs.~\ref{Fig:yuknh} and ~\ref{Fig:yukih}, the relations between the leptoquark Yukawa couplings $y_S^{ij}$ and mass $m_S$ are shown for survived sample points in NH and IH respectively. We see that the LFV processes set more strict bounds than the FCNC ones. And fewer points could interpret the $R_{K^{(*)}}$ anomaly which restricts the Wilson coefficients $C_{9}^e-C_{9}^\mu$ to a pretty narrow region.

We find a few features in the Yukawa couplings which could interpret the $R_{K^{(*)}}$ anomaly while satisfying all other constraints. There exists a hierarchy between $|y_S^{i1}|$ and $|y_S^{i2(3)}|$ in that $|y_S^{i1}|$ is always smaller than $|y_S^{i2(3)}|$ for both cases of NH and IH. All of $|y_S^{i1}|$ reach the largest value around $0.01$ in NH except for a few points exceeding the value for $m_{S}>5~\TeV$. But in the case of IH, $|y_S^{21}|$ and $|y_S^{31}|$ are always smaller than $0.01$ while $|y_S^{11}|$ can reach a large value of $0.1$ at some points. The relatively smaller values of $|y_S^{i1}|$ are mainly due to the strict constraint on the $\mu-e$ conversion which relates to $|y_S^{11}|$ and $|y_S^{21}|$. In the meanwhile, $|y_S^{i2}|$ and $|y_S^{i3}|$ are close to being degenerate except for $|y_S^{32}|$ and $|y_S^{33}|$ in the case of IH. Finally, the allowed values of $y_S$s increase with $m_{S}$ because the actual constraints are set on the combination $y_S^2/m_{S}^2$.

\begin{figure}
	\centering
	\includegraphics[width=1\linewidth]{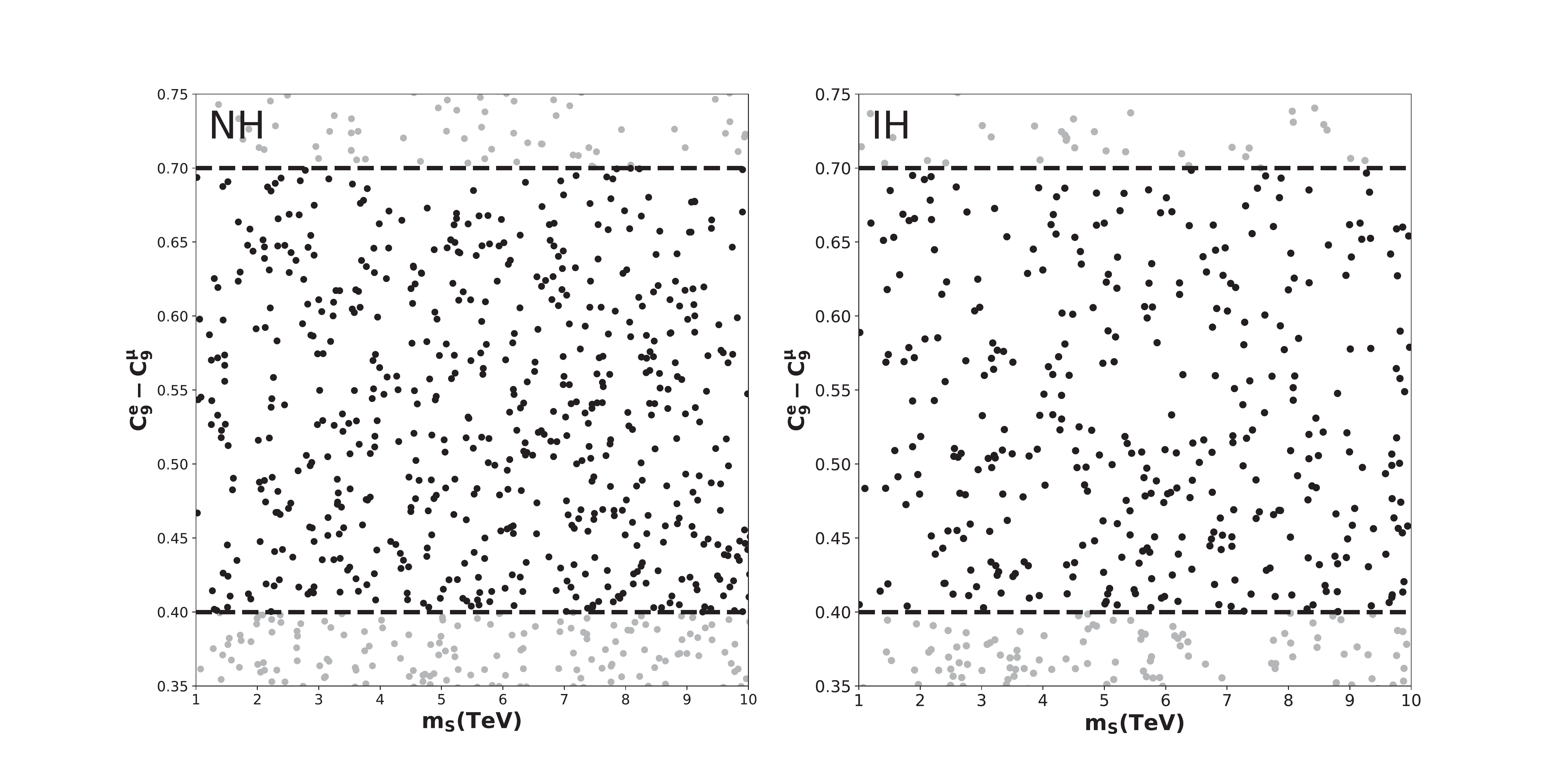}\\
	\caption{The distribution of $C_9^e - C_9^\mu$ for NH (left panel) and IH (right). The region between the two dashed lines is the $1\sigma$ range favored by the $R_{K^{(*)}}$ anomaly.}
	\label{Fig:C9distr}
\end{figure}

After imposing all the constraints in Sec.~\ref{CT}, we get the survived regions of parameters. In Fig.~\ref{Fig:C9distr} we present the distribution of sample points for $C_9^e - C_{9}^\mu$ in the cases of NH and IH as a function of the leptoquark mass $m_S$. While the anomaly can be accommodated for both NH and IH, the case of NH has more survived samples.  In Fig.~\ref{Fig:P5}, we show the individual contributions of $C_9^{e,\mu}$ to the $R_{K^{(*)}}$ anomaly. It is clear that for NH the dominant contribution is $C_9^\mu$ while $C_9^e\approx0$. For IH however, both $C_9^e$ and $C_9^\mu$ are important. The model could also accommodate the potential $P_5^\prime$ anomaly in $B\to K^*\mu^+\mu^-$ and angular anomaly in $B_s^0\to \phi \mu^+\mu^-$, both of which only require a new contribution in $b\to s \mu^+\mu^-$, with a favored $1\sigma$ range $C_9^\mu\in[-0.76,-0.48]$~\cite{Alok:2017sui}.

\begin{figure}
    \includegraphics[width=1\linewidth]{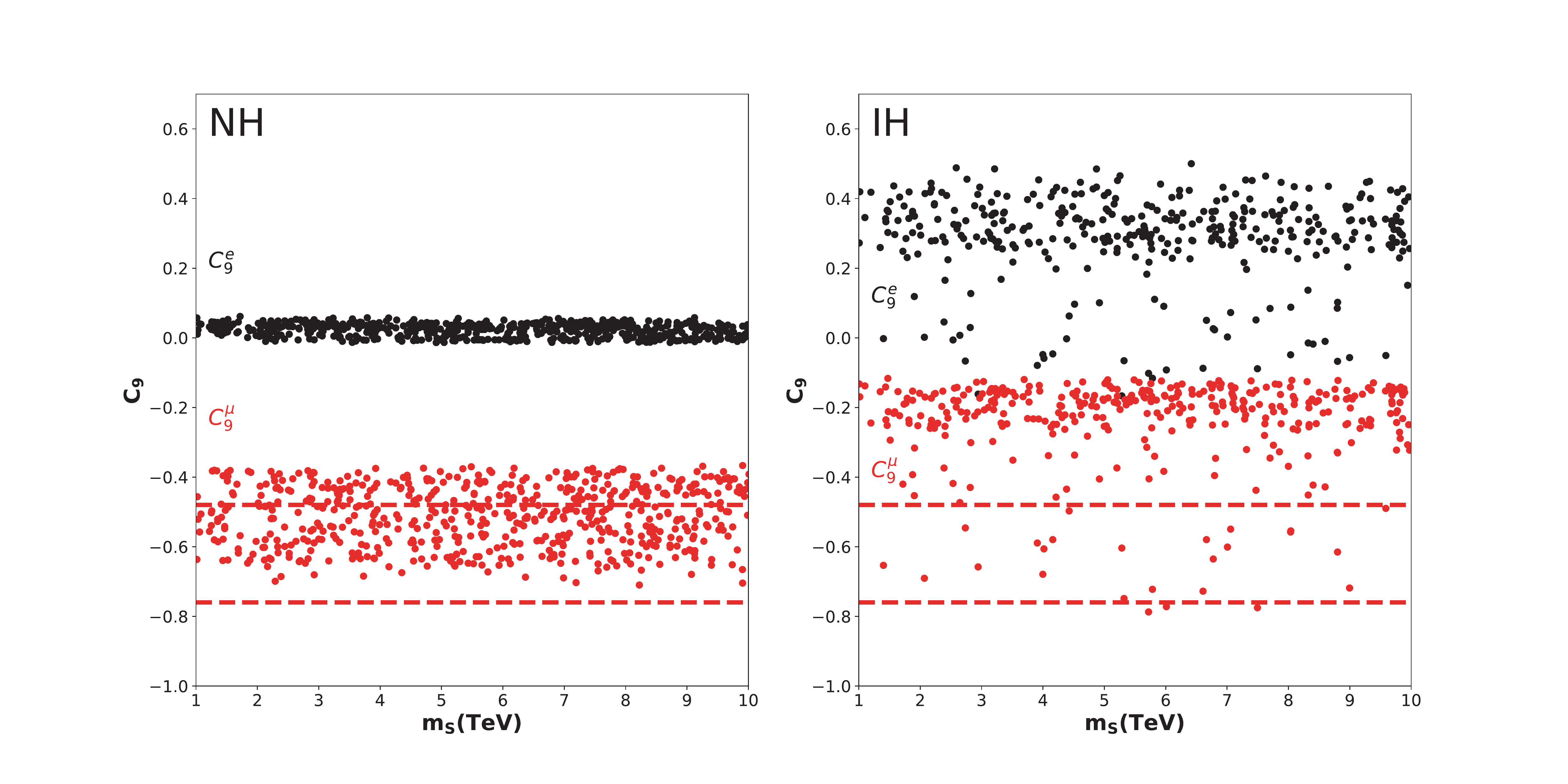}
    \caption{The distributions of $C_9^{e,\mu}$ for NH (left panel) and IH (right). The region between the two dashed lines is the $1\sigma$ range favored by the potential $P_5^\prime$ anomaly and anomaly in $B_s^0\to \phi \mu^+\mu^-$~\cite{Alok:2017sui}.}
    \label{Fig:P5}
\end{figure}

Inspired by Refs.~\cite{Glashow:2014iga,Boucenna:2015raa,Crivellin:2015era,Chiang:2017hlj,
Crivellin:2017dsk}, we further examine the LFV decays of the $B$ meson. Currently, the most stringent constraints come from the $B^+$ meson decays~\cite{Olive:2016xmw}:
\begin{align}\label{BRCT}
&\text{Br}(B^+ \to K^+ \mu^\pm e^\mp) < 9.1\times 10^{-8}, \\
&\text{Br}(B^+ \to K^+ \mu^\pm \tau^\mp) < 4.8\times 10^{-5}, \\
&\text{Br}(B^+ \to K^+ \tau^\pm e^\mp) < 3.0\times 10^{-5}, \label{BRCT1}
\end{align}
where a sum over oppositely charged leptons is implied. The branching ratios can be normalized to that for the lepton flavor conserving decay $B^+ \to K^+ \mu^+ \mu^-$~\cite{Glashow:2014iga,Crivellin:2015era}:
\begin{equation}
	\frac{\text{Br} (B^+ \to K^+ \ell_i^\pm \ell_j^\mp)}{\text{Br} (B^+ \to K^+ \mu^+ \mu^-)} = \phi_{ij} \frac{|(y_S^{i3}) (y_S^{j2*})|^2 + |(y_S^{j3}) (y_S^{i2*})|^2}{ |4 m_S^2\mathcal{A}_{\text{SM}} + (y_S^{23}) (y_S^{22*})|^2},
\end{equation}
where $\ell_i\ne\ell_j$ and $\mathcal{A}_{\text{SM}}$ denotes the SM amplitude dominated by $C_9^{\text{SM}}=-C_{10}^{\text{SM}}$ while ignoring the smaller $C_7^{\text{SM}}$ term~
{\footnote{{The incurred error can be estimated as follows. The transition matrix element of the operator $\bar s\sigma_{\mu\nu}P_Rbq^\nu$ between the mesons $B(p_B)$ and $K(p_K)$ is parameterized by a dimensionless form factor associated with the structure, $q^2(p_B+p_K)_\mu-(p_B^2-p_K^2)q_\mu$, with $q=p_B-p_K$ and $p_{B,K}^2=m_{B,K}^2$, see, e.g., \cite{Isgur:1990kf}. Therefore in the amplitude for $B\to K\mu^+\mu^-$, the $q_\mu$ term due to $C_7^{\text{SM}}\mathcal{O}_7$ vanishes by current conservation while the photon propagator is removed by the explicit factor of $q^2$ in the first term. The latter combines with the $C_9^{\text{SM}}\mathcal{O}_9$ term, with relative importance measured by $[C_7^{\text{SM}}/C_9^{\text{SM}}][2m_b/(m_B+m_K)]\sim 0.11$ times an order one ratio of the two form factors~\cite{latticeandLCSR}. Including the separate contribution to the decay rate from the $C_{10}^{\text{SM}}\mathcal{O}_{10}$ term, dropping the $C_7^{\text{SM}}\mathcal{O}_7$ term causes an error of about $12\%$ in the decay rate.}}}
,
\begin{equation}
	\mathcal{A}_{\text{SM}} = - \frac{4 G_F}{\sqrt{2}} V_{tb}^* V_{ts} \frac{\alpha}{4 \pi} C_9^{\text{SM}},
\end{equation}
with the Wilson coefficient $C_9^{\text{SM}}(m_b) = 4.2$~\cite{Hiller:2014yaa}, and $\phi_{e\mu}\simeq1,~\phi_{e\tau}=\phi_{\mu\tau}\simeq0.63$ account for phase space differences~\cite{Boucenna:2015raa}. The LHCb collaboration has yielded $\text{Br}(B^+ \to K^+ \mu^+ \mu^-) = 4.29 \times 10^{-7}$~\cite{Aaij:2014pli}. The distributions of branching ratios are presented in Fig.~\ref{Fig:brsLFV}. As one can see clearly, all LFV decay branching ratios are under the current limits. While for IH the sample points are concentrated for each decay, the points are more scattered in the case of NH. The largest LFV branching ratio under NH arises in the channel $B^+ \to K^+ \mu^\pm \tau^\mp$ and is about $10^{-9}$, while under IH the branching ratios for the other two channels are the largest but of order $10^{-11}$. We thus observe a correlation between the neutrino mass hierarchy and the LFV decays of the $B^+$ meson:
\begin{eqnarray}\label{RL}
\text{Br}(B^+ \to K^+ \mu^\pm \tau^\mp)\gtrsim\text{Br}(B^+ \to K^+ \mu^\pm e^\mp)\approx\text{Br}(B^+ \to K^+ \tau^\pm e^\mp)~\text{for NH},\\
\text{Br}(B^+ \to K^+ \mu^\pm \tau^\mp)\ll\text{Br}(B^+ \to K^+ \mu^\pm e^\mp)\approx\text{Br}(B^+ \to K^+ \tau^\pm e^\mp)~\text{for IH}.~~
\end{eqnarray}
From the experimental point of view, the decay $B^+ \to K^+ \mu^\pm e^\mp$ is the easiest to search for. Considering the LHCb RUN II will reach the region of $\mathcal{O}(10^{-10})$ in this channel \cite{Boucenna:2015raa}, only NH will have a chance to be detected in certain parameter regions.

\begin{figure}
	\centering
	\includegraphics[width=1\linewidth]{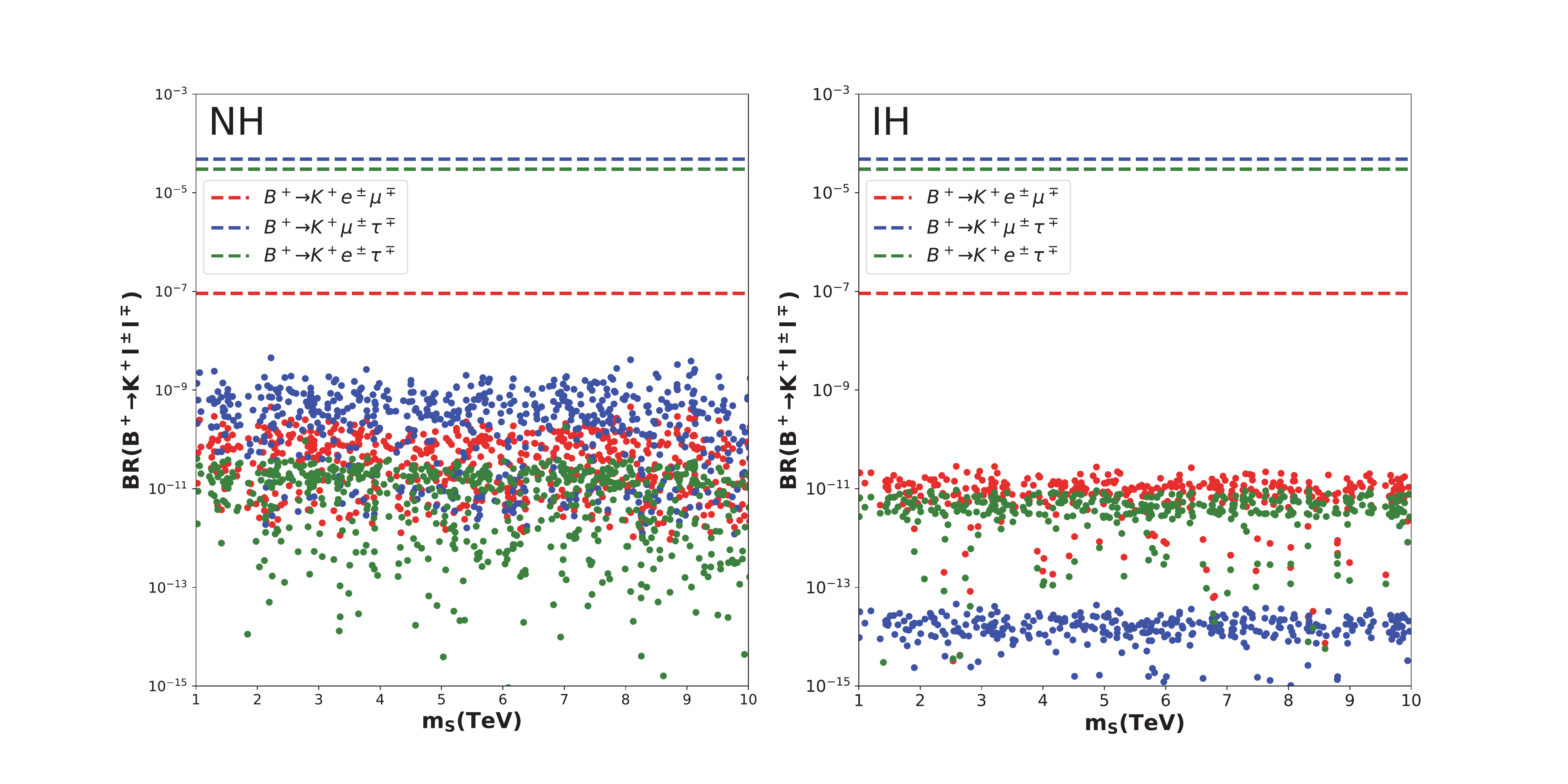}
	\caption{The distributions of LFV decay branching ratios for NH (left panel) and IH (right). The three dashed lines represent the current upper limits in Eqs.~(\ref{BRCT})--(\ref{BRCT1}).}
	\label{Fig:brsLFV}
\end{figure}

\section{Conclusion}\label{CL}

We have studied the $R_{K^{(*)}}$ anomaly in the colored Zee-Babu model that accounts for tiny neutrino mass as a two-loop effect due to interactions with a leptoquark and a diquark. Upon parameterizing the $R_{K^{(*)}}$ anomaly related Yukawa coupling $y_S$ in terms of neutrino parameters and free parameters, we successfully acquired parameter regions that can accommodate the anomaly while evading various constraints from LFV and FCNC processes and direct searches for leptoquarks and diquarks. We found that the anomaly can be interpreted for both NH and IH scenarios. For the phenomenologically viable sample points in parameter space, there is a hierarchical structure in the leptoquark Yukawa couplings $y_S$, e.g., $|y_S^{i1}|\ll|y_S^{i2(3)}|$. For the first time, we have examined the LFV decays of the $B^+$ meson in a complete neutrino mass model with a leptoquark, and revealed a strong correlation between neutrino mass hierarchies and its dominant LFV decays. Experimentally speaking, only the decay mode $B^+ \to K^+ \mu^\pm e^\mp$ is promising to be probed at the LHCb RUN II in certain parameter regions in the case of NH.

\section*{Acknowledgement}

This work was supported in part by the Grants No. NSFC-11575089 and No. NSFC-11025525, by The National Key Research and Development Program of China under Grant NO. 2017YFA0402200, and by the CAS Center for Excellence in Particle Physics (CCEPP). We thank the anonymous referee for comments that have helped us improve our presentation and made our statements more precise.


\end{document}